\documentclass[twocolumn,showpacs,preprintnumbers,amsmath,amssymb,%
prd,floatfix,nofootinbib,superscriptaddress]{revtex4-1}  

\usepackage{graphicx}
\usepackage{epsfig}
\usepackage{color}
\usepackage{longtable}
\usepackage{hyperref}
\usepackage{breakurl}
\usepackage{latexsym}
\usepackage{amsmath}
\usepackage{amssymb}
\usepackage{dsfont}
\usepackage{bm}

\newcommand{\ZE}{\mathbb{Z}}
\newcommand{\I}{\mathrm{i}}  
\newcommand{\E}{\mathrm{e}}  
\newcommand{\eq}[1]{(\ref{#1})}
\newcommand{\FC}{\;,}
\newcommand{\FD}{\;.}
\newcommand{\Tr}{\mathrm{Tr}}
\begin{document}


\date{\today}
\title{Coupled channel analysis of the $\rho$ meson decay in lattice QCD}

\vspace{0.2cm}

\author{C. B. Lang}
\email{christian.lang@uni-graz.at}
\affiliation{Institut f\"ur Physik, FB Theoretische Physik, Universit\"at
Graz, A--8010 Graz, Austria}

\author{Daniel Mohler}
\email{mohler@triumf.ca}
\affiliation{TRIUMF, 4004 Wesbrook Mall Vancouver, BC V6T 2A3, Canada}

\author{Sasa Prelovsek}
\email{sasa.prelovsek@ijs.si}
\affiliation{Department of Physics, University of Ljubljana, Slovenia}
\affiliation{Jozef Stefan Institute, Ljubljana, Slovenia}

\author{Matija Vidmar}
\affiliation{Jozef Stefan Institute, Ljubljana, Slovenia}

\begin{abstract}
We employ a variational basis with a number of $\bar{q}q$ and $\pi\pi$ lattice
interpolating fields with quantum numbers  of the $\rho$ resonance to extract
the discrete energy spectrum in a finite volume. In the elastic region, this
spectrum is related to the phase-shift of the continuum scattering amplitude by
L\"uscher's formula and the relation allows the extraction of resonance
parameters from the spectrum calculation.  The simulations are performed at
three different total momenta of the coupled $\bar q q-\pi\pi$ system, which
allows us to extract the p-wave scattering phase at five values of pion relative
momenta near the resonance region. The effective range formula describes the
phase-shift dependence nicely and we extract the resonance mass
$m_\rho=792(7)(8)$~MeV and the coupling $g_{\rho\pi\pi}=5.13(20)$ at our
$m_\pi\simeq 266~$MeV. The coupling  $g_{\rho\pi\pi}$  is directly related to 
the width of the $\rho$ meson and our value is close to the value derived from
the experimental width. The simulations are performed using dynamical gauge
configurations with two mass-degenerate flavors of tree-level improved
clover-Wilson fermions. Correlation functions are calculated using the recently
proposed distillation method with Laplacian  Heaviside (LapH) smearing of
quarks, which enables flexible calculations, in many cases with
unprecedented accuracy.

\end{abstract}

\maketitle 


\section{Motivation and Introduction}

\textbf{This version contains an erratum at the end. The main text is unchanged
and identical to arXiv:1105.5636v2.}
Almost all hadrons listed in the Particle Data Group \cite{Nakamura:2010zzi} tables are unstable, most of them decaying strongly. In quenched calculations, where vacuum
quark loops are  disregarded, all hadronic states appear as stable states. In
full QCD, on the other hand,  truly asymptotic exponential behavior is always dominated by the
lowest stable end product. 
This is unsatisfactory.

In continuum physics experiments resonances are identified via the scattering
cross section and subsequent phase-shift analyses.
 In the lattice discretization
of QCD, instead, one studies the correlation functions of hadron interpolators
for Euclidean time distances. The result is a combination of exponentially
decaying terms, each corresponding to the energy level of a contributing
eigenstate. Due to the finiteness of the lattice system, the energy levels are
discrete. The spectral density is related to a discretization of the cross
section. However, in realistic lattice simulations only very few such levels can
be determined. The typical gaps are $\mathcal{O}(2\pi/L)$ for lattices of
spatial extent $L$; for most simulations this corresponds to level spacing
$\mathcal{O}(400)$ MeV.

However, as has been pointed out in a seminal paper by L{\"u}scher
\cite{Luscher:1986pf,Luscher:1990ux}, for a resonating system  the discrete
spectrum obtained in a finite volume can be related to the phase-shift of the
continuum scattering amplitude in the elastic  region. The resulting volume dependence of the
spectrum can then be used to explore the resonance properties
\cite{Luscher:1991cf}. Model simulations in two dimensions \cite{Gattringer:1992yz} as well
as in four dimensions  \cite{Gockeler:1994rx} demonstrated the feasibility of that approach.
The original derivation in the decaying particles rest frame was then extended
to moving frames  \cite{Rummukainen:1995vs,Kim:2005zzb,Feng:2011ah},  thus
enhancing  the practical applicability, allowing one to obtain the phase-shift at
more momentum points  for a given lattice size. Because of several problems
there have only been a few attempts to apply that scheme to the decay
$\rho\to\pi\pi$
\cite{Aoki:2007rd,Gockeler:2008kc,Aoki:2010hn,Feng:2010es,Feng:2011ah,
Frison:2010ws}, while the first lattice estimate of the $\rho\rightarrow\pi\pi$ amplitude \cite{McNeileMichael} did not 
apply L\"uscher's method. Note that widths for most of the other resonances
have not been determined on the lattice at all. 

There are two major complications. The first one concerns the hadronic
lattice interpolators used. Let us assume  that we work with the fully dynamic
vacuum, i.e., including the dynamical quark vacuum loops in a full QCD 
simulation. Naively one would expect that, even if one correlates only
quark-antiquark interpolators with the correct quantum numbers of the $\rho$,
due to the vacuum loops, $\pi\pi$ intermediate states should also contribute and
affect the energy levels accordingly. This is hardly observed; actually already
in model calculations \cite{Gattringer:1992yz} it proved necessary to include both,
the heavy boson and the two light bosons in the set of interpolators. Similar
observations were made in other calculations involving baryon and meson
correlation functions \cite{Engel:2010my,Bulava:2010yg,Morningstar:2011ka,Dudek:2010wm}. The
obvious interpretation is that the overlap of the quark-antiquark
interpolators with the  meson-meson decay channel interpolators is too weak to
have been observed.

For that reason one should extend the set of hadron interpolators to include
both, various versions of the quark-antiquark interpolator (like, e.g.,
different Dirac structure or different quark smearing functions), as well as 
meson-meson interpolators. The latter involve four propagating fermions 
and the corresponding
entries of the correlation function  usually will involve backtracking loops. In
addition to this technical complication there is also the notorious issue of
statistical weight for such contributions. The so-called distillation  (or
Laplacian-Heaviside quark smearing) method introduced in 
\cite{Peardon:2009gh} and employed 
in \cite{Dudek:2009qf,Dudek:2010wm,Dudek:2010ew,Dudek:2011tt,Edwards:2011jj}
helps us significantly to deal with that problem.

The second challenge concerns the energy levels. One works with several
hadronic interpolators, all with the  correct quantum numbers and total momentum
in the given channel. The diagonalization of the correlation matrix  gives the
eigenstates and eigenenergies according to the so-called variational method
\cite{Michael:1985ne,Luscher:1985dn,Luscher:1990ck,Blossier:2009kd}. The set of
lattice interpolators should be large enough to be able to represent the
leading eigenstates and thus the leading energy levels. The better the set is,
the better the results will be and the more energy levels can be determined,
depending of course also on the available statistics. 
In previous calculations aimed at $\rho$ meson decay, at most two 
interpolators were used: one quark-antiquark and one pion-pion interpolator. 
We extend this to a larger interpolator basis.

For our calculation we use one lattice ensemble with $n_f=2$ dynamical mass-degenerate light quarks and clover-improved Wilson fermionic action (generated in
context of the work \cite{Hasenfratz:2008ce,Hasenfratz:2008fg} in order to study
reweighting techniques). The ensemble consists of $16^3\times 32$ lattices with spatial extent 1.98
fm and $m_\pi \simeq 266~$MeV. We consider cross-correlations of several
interpolators (16 for the $\rho$ channel, 6 for the pion channel) and solve the
generalized eigenvalue problem to reliably determine the two lowest energy
levels. We study the $\rho$ channel for three values of the total momentum and
obtain the elastic phase-shift in the resonance region.

Section \ref{Tools} gives an overview of the methods:  quarks sources,
interpolators, variational analysis, phase-shift relations and finite time effects. In Sect.
\ref{Computations} the set of configurations and details on the computations are
summarized and in Sect. \ref{Results} we discuss the results: correlation
functions, energy levels, phase-shift and resonance parameters. 

Reference \cite{Bernard:2008ax} suggests an alternative approach which has
recently been investigated in \cite{Giudice:2010ch}. Furthermore another
procedure has been suggested in \cite{Meissner:2010rq}.

\section{Tools}\label{Tools}

\subsection{Phase-shift formulas, brief review}\label{sec:tools_phase_shifts}

On finite lattices there are, strictly speaking, no asymptotically free states
and the energy spectrum is always discrete. It was pointed out by L{\"u}scher
\cite{Luscher:1990ux,Luscher:1991cf} that, assuming a localized interaction
range, the energy level of a correlation matrix for channels with resonances in
a finite volume can be related to the corresponding phase-shift in infinite
volume  in the elastic region (i.e., where only one decay channel is open). The
relation was derived for interpolators with spatial momentum zero. For a
particle like the $\rho$ meson, which can decay into two pions with back-to-back
momenta, the available momenta are discrete on finite lattices and depend on the
spatial extent.

In the noninteracting case the various two-pion energy levels
will decrease with growing volume and this leads to level crossing with the
stable $\rho$ state. If interaction is switched on the level crossing is avoided
and the energy levels ``change their identity''. This was demonstrated in a two
dimensional
resonance model in \cite{Gattringer:1992yz} as well as in four dimensional $\phi^4$-model
simulations \cite{Gockeler:1994rx}.

For the analysis of resonances in that method one needs several ingredients. The
set of interpolators should overlap with both, the single particle content
(i.e., for a meson the quark-antiquark component) as well  as the two particle
content (i.e., the meson-meson decay channel). Furthermore it should be possible
to analyze more levels than just the ground state energy. Third, in the
originally proposed method one needs several spatial volumes to obtain the phase
shift at several values of relative momentum. This makes the approach costly.

The third aspect can be ameliorated, though, by studying also channels with
nonvanishing total momentum 
\begin{equation}
\label{eq_def_P}
\mathbf{P}=\frac{2\pi}{L}\mathbf{d}\quad\textrm{with}\quad\mathbf{d}\in\ZE^3\FD
\end{equation} 
In our simulation we study the cases
\begin{equation}
\label{eq_def_d}
 \mathbf{d}=(0,0,0),\ (0,0,1),\ (1,1,0)
\end{equation}
and permutations, which have previously been combined in the simulation \cite{Feng:2010es}. 
 Different values of $\mathbf{P}$ allow to  obtain the phase
shifts at different values of pion relative momenta. The lowest $\pi\pi$ state
in the $\rho$ channel with $|\mathbf{P}|=0$ is  $\pi(2\pi/L)\;\pi(-2\pi/L)$ (due to
$\ell=1$) and is significantly  above the $\rho$ resonance in typical simulations.
In the case of a $\rho$ with  $|\mathbf{P}|=2\pi/L$, the  $\pi(0)\;\pi(2\pi/L)$ is
closer to the resonance region, for example.  However, this case involves relativistic
kinematics in the nonzero momentum frame  as  pointed out in
\cite{Rummukainen:1995vs,Kim:2005zzb,Feng:2011ah}.  The relativistic distortion reduces
the full cubic symmetry $O_h$ to that of prismatic dihedral groups, i.e., to the
symmetry of a cuboid (quadratic prism) $D_{4h}$   for total momenta of type
(0,0,1) and to the symmetry of a rhombic prism  $D_{2h}$ for momentum (1,1,0).

In the laboratory frame, the total 3-momentum of two {\it noninteracting} bosons in a cubic lattice of volume $L^3$ and periodic boundary conditions is
\begin{equation}
\mathbf{P}=\mathbf{p}_1+\mathbf{p}_2= \frac{2\pi}{L}\mathbf{d}\quad 
\end{equation}
and the energy is
\begin{eqnarray}
E=E_1+E_2&=&\sqrt{m^2+\mathbf{p}_1^2}+\sqrt{m^2+\mathbf{p}_2^2}\nonumber\\
\quad\textrm{with}\quad\mathbf{p}_i&=&\frac{2\pi}{L}\mathbf{n}_i\;,\;
\mathbf{n}_i\in\ZE^3\FD
\end{eqnarray}
The velocity $\mathbf{v}=\mathbf{P}/E$ gives the relativistic boost  factor
$\gamma=1/\sqrt{1-\mathbf{v}^2}$. 
In the center-of-momentum frame (CMF) the total momentum vanishes and the 
bosons momenta are 
\begin{equation}
\mathbf{p}_1^*=-\mathbf{p}_2^*\equiv \mathbf{p}^*\FD
\end{equation}
The energy in the CMF is
\begin{equation}
E_{CM}=2\sqrt{m^2+\mathbf{p}^{*2}}=E/\gamma\FC
\end{equation}
and the momentum is related to the laboratory frame through
\begin{equation}
\mathbf{p}^{*}=\frac{1}{2} \mathbf{\gamma}^{-1}_{op}(\mathbf{p}_1-\mathbf{p}_2)\FC
\end{equation}
where the boost factor acts in direction of $\mathbf{v}$,
\begin{equation}
 \mathbf{\gamma}^{-1}_{op}  \mathbf{p}\equiv \mathbf{p}_\parallel/\gamma+ 
 \mathbf{p}_\perp\FC\;\; 
 \mathbf{p}_\parallel= \mathbf{v} ( \mathbf{p}\cdot \mathbf{v})/| 
 \mathbf{v}|^2\FC\;\; 
 \mathbf{p}_\perp= \mathbf{p}- \mathbf{p}_\parallel\FD
\end{equation}
The relativistic 4-momentum squared is invariant, thus the relation to the
laboratory energy $E$ is
\begin{equation}\label{ecm}
E_{CM}^2=E^2-\mathbf{P}^2\quad\rightarrow\quad \mathbf{p}^{*2}=
\frac{1}{4} E_{CM}^2-m^2\FD
\end{equation}
Due to the coarseness of the lattice we replace in our calculations this
continuum dispersion relation by the lattice dispersions relation as suggested
in \cite{Rummukainen:1995vs}, i.e., 
\begin{eqnarray}
\label{dispersion_lat}
\cosh{E_{CM}a}&=&\cosh{Ea}-2 \sum_{k=1}^3 \sin^2\left(\frac{P_ka}{2}\right)\;,\\
\left(2 \sin{\frac{a\,p^*}{2}}\right)^2&=&2\cosh{\frac{E_{CM}a}{2}}-2\cosh{ma}\FD
\end{eqnarray}

For the \emph{interacting} case, the momenta $p_{1,2}$ of individual pions 
in the laboratory frame are no longer multiples of
$2\pi/L$. Assuming a localized interaction region one associates the outside
region with that of two free bosons. The observed energy levels  $E_n$ are
shifted and related to the scattering phase-shift. Expressed through the CMF variable 
\begin{equation}\label{pstar}
\mathbf{p}^{*2}\equiv \left(q\frac{2\pi}{L}\right)^2\FC
\end{equation}
one obtains relations of the form $\tan \delta(q)=f(q)$ for transcendental
functions $f(q)$.

We concentrate on the decay $\rho\to\pi\pi$ where the two pions are in p-wave
($\ell=1$). Details have been discussed in the original papers
\cite{Luscher:1990ux,Luscher:1991cf,Rummukainen:1995vs,Kim:2005zzb,Feng:2010es,Feng:2011ah}. For completeness
we summarize here only the relevant final expressions, where phase-shifts 
are expressed in terms of the generalized zeta function defined by 
\begin{align}
\label{zeta}
\mathcal{Z}_{\ell m}^{\mathbf{d}}(s;q^2) 
&=
\sum_{\mathbf{x}\in P_\mathbf{d}} 
\frac{\mathcal{Y}_{\ell m}^*(\mathbf{x})}{(\mathbf{x}^2-q^2)^s}\FC\\
P_\mathbf{d}
&=
\left\{\mathbf{x}\in \mathbb{R}^3\mid \mathbf{x}=\mathbf{\gamma}^{-1}_{op}
\left(\mathbf{m}+\frac{\mathbf{d}}{2}\right),\; \mathbf{m}\in \ZE^3\right\}\FC\nonumber\\
\mathcal{Y}_{\ell m}(\mathbf{x})
&=
|\mathbf{x}|^\ell Y_{\ell m}(\mathbf{x})\FC\nonumber
\end{align}
and $\mathcal{Y}_{\ell m}$ are the harmonic polynomials to the spherical
harmonics functions $Y_{\ell m}$.
The zeta function has to be analytically continued to $s=1$. The simpler form
for $\mathbf{d} =0$ is given in \cite{Luscher:1990ux}. A rapidly convergent
expression for nonvanishing $\mathbf{d}$ is derived in \cite{Feng:2011ah}. We
numerically compared the different representations of the zeta functions of
\cite{Kim:2005zzb} and \cite{Feng:2011ah} and found agreement.

The symmetry groups of the sum appearing in $Z_{lm}$ (\ref{zeta}) are $O_h$, $D_{4h}$ and $D_{2h}$ respectively for $d=(0,0,0),~(0,0,1)$ and $(1,1,0)$. The $J^{P}=1^{-}$ states appear in the specific representations of these symmetry groups and the final expressions for the phase-shifts are:  
\hspace{3pt}\\

\noindent {\bf Zero momentum} $\mathbf{P=(0,0,0)}$  \\(for irrep $T^-_1$ in $O_h$) \cite{Luscher:1990ux}:
\begin{equation}
\tan \delta(q)=\frac{\pi^{3/2} q}{\mathcal{Z}_{00}(1;q^2)}\FD
\end{equation}

\hspace{3pt}\\

\noindent  {\bf Nonzero momentum} $\mathbf{P=(0,0,1)\tfrac{2\pi}{L}}$ \\(for irrep
$A^-_2$ in $D_{4h}$) \cite{Rummukainen:1995vs}:
\begin{equation}
\tan \delta(q)=\frac{\gamma \pi^{3/2} q^3}{q^2 
\mathcal{Z}_{00}^{\mathbf{d}}(1;q^2)
+\sqrt{\frac{4}{5}}\;\mathcal{Z}_{20}^{\mathbf{d}}(1;q^2)}\FD
\end{equation}

\hspace{3pt}\\

\begin{widetext}
\noindent {\bf Nonzero momentum} $\mathbf{P=(1,1,0)\tfrac{2\pi}{L}}$ \\(for irrep
$B^-_1$ in $D_{2h}$) \cite{Feng:2010es}:
\begin{equation}
\tan \delta(q)=\frac{\gamma \pi^{3/2} q^3}{q^2 
\mathcal{Z}_{00}^{\mathbf{d}}(1;q^2)
-\sqrt{\frac{1}{5}}\;\mathcal{Z}_{20}^{\mathbf{d}}(1;q^2)
+\I\sqrt{\frac{3}{10}}\;(\mathcal{Z}_{22}^{\mathbf{d}}(1;q^2)-
\mathcal{Z}_{2\bar2}^{\mathbf{d}}(1;q^2))}\FD
\end{equation}
We independently derived this relation and we agree with this expression, originally  presented in  \cite{Feng:2010es,Feng:2011ah}.
\end{widetext}

\subsection{Variational analysis}

To extract the lowest two energy levels  with the quantum numbers
$I^G(J^{PC})=1^+(1^{--})$ 
of the $\rho$ meson as well as the ground state energies with quantum numbers $I^G(J^{PC})=1^-(0^{-+})$ of the pion, 
we construct a matrix
$C(t)_{ij}$ of lattice interpolating fields containing both quark-antiquark and
meson-meson (in our case pion-pion) interpolators
\begin{equation}
C(t)_{ij}=\sum_n\mathrm{e}^{-t\,E_n}\big\langle 0|O_i|n\big\rangle\big\langle
 n|O_j^\dagger|0 \big\rangle .
\end{equation}
For this matrix, the generalized eigenvalue problem
\begin{equation}
\label{variational}
C(t)\vec{\psi}^{(n)}=\lambda^{(n)}(t)C(t_0)\vec{\psi}^{(n)}
\end{equation}
is solved for each time slice. For the eigenvalues $\lambda^{(n)}(t)$ one 
obtains
\begin{equation}
\lambda^{(n)}(t)\propto\E^{-t\,E_n}\left(1+\mathcal{O}
\left(\E^{-t\,\Delta E_n}\right)\right) ,
\end{equation}
so that each eigenvalue is dominated by a single energy at large time
separations. This method is called the variational method
\cite{Luscher:1990ck,Michael:1985ne,Luscher:1985dn,Blossier:2009kd}. For a detailed discussion
of the energy difference $\Delta E_n$, which is in general given by the
difference between the energy level in consideration and the closest
neighboring level, please refer to \cite{Blossier:2009kd}.

We calculate the eigenvector components of the regular eigenvector problem
\begin{align}
C(t_0)^{-\frac{1}{2}}C(t)C(t_0)^{-\frac{1}{2}}\vec{\psi}^{(n)\,\prime}
&=\lambda^{(n)}(t)\vec{\psi}^{(n)\,\prime}\FD
\end{align}
In addition to the
eigenvalues, the eigenvectors provide useful information and can serve as a
fingerprint for a given state. To track the eigenvalue corresponding to a given
energy over the full range of time separations, the eigenvalues have to be
sorted, either by their magnitude or by scalar products of their eigenvectors.
In the presence of backwards running contributions caused by the finite time
extent of the lattice, a combination of both methods works well:  the eigenvalues are
sorted by magnitude at low time separations and by scalar products at larger
time separation. For our analysis we choose this method.

\subsection{Interpolators}

For the $\rho$ channel
we employ fifteen quark-antiquark interpolators and one pion-pion interpolator
with $J^{PC}=1^{--}$ and  $|I,I_3 \rangle =|1,0\rangle $ in the variational
basis for each of the three choices for $\mathbf{P}$ as given in \eq{eq_def_d}. All previous
simulations aimed at determining the $\rho$ meson width used at most one 
quark-antiquark and
one pion-pion interpolator and extracted the two lowest energy levels from a $2\times
2$ variational basis. This may not be reliable if the third energy level is
nearby and does not allow testing whether the resulting two levels are robust 
against the choice of interpolators. A larger basis enables us to exploit
the  dependence of the extracted energies on the choice of the interpolators. It
also indicates  whether the lowest two states can  be reliably extracted
using our quark-antiquark interpolators alone, or whether 
the pion-pion interpolators are required in the variational basis. 

The 15 different quark-antiquark interpolators  ${\cal
O}_{type}^{s}$ ($type=1,..,5,\ s=n,m,w$)  differ in  type (Dirac and color structure)
and width of the smeared quarks $q_s$.
We use three different smearing widths $s=n,\,m,\,w$
(narrow, middle, wide) for  individual quarks and all quarks  in a given
interpolator have the same width $s$ in this simulation.
(Choosing
different quark widths within  an interpolator is a straightforward
generalization and one just needs to pay attention that the resulting 
$C$-parity is correct.)
The details on the smearing are given in Subsect. \ref{sec:smearing}. 
The interpolator ${\cal O}_{6}$ is the $\pi\pi$ interpolator whose structure is
explained at the end of this subsection. Our sixteen $\rho$ interpolators are:
\begin{widetext}
\begin{align}
\label{interpolators}
{\cal O}_{1}^{s}(t)&=\sum _{\mathbf{x},i}\tfrac{1}{\sqrt{2}}~\bar u_s(x) ~
A_i\gamma_i~ \E^{\I \mathbf{Px}} ~u_s(x)\ -\{u_s\leftrightarrow d_s\}\qquad 
(s=n,m,w) \FC\nonumber\\
{\cal O}_{2}^{s}(t)&=\sum_{\mathbf{x},i}\tfrac{1}{\sqrt{2}}~\bar u_s(x)~
\gamma_t  A_i\gamma_i~ \E^{\I \mathbf{Px}}~ u_s(x)\ -\{u_s\leftrightarrow d_s\} 
\qquad(s=n,m,w) \FC\nonumber\\
{\cal O}_{3}^{s}(t)&=\sum_{\mathbf{x},i,j}\tfrac{1}{\sqrt{2}}~\bar u_s(x)
\overleftarrow{\nabla}_j~A_i\gamma_i~ \E^{\I \mathbf{Px}}~
\overrightarrow{\nabla}_j u_s(x)\ - \{u_s\leftrightarrow d_s\}\qquad(s=n,m,w)\FC\nonumber\\
{\cal O}_{4}^{s}(t)&=\sum_{\mathbf{x},i}\tfrac{1}{\sqrt{2}}~\bar u_s(x)~A_i~
\tfrac{1}{2} [\E^{\I \mathbf{Px}}~
\overrightarrow{\nabla}_i -\overleftarrow{\nabla}_i 
\E^{\I \mathbf{Px}}]u_s(x) \ -\{u_s\leftrightarrow d_s\}
\qquad(s=n,m,w)\FC\nonumber\\
{\cal O}_{5}^{s}(t)&=\sum_{\mathbf{x},i,j,k} \tfrac{1}{\sqrt{2}}~\epsilon_{ijl} 
 ~\bar u_s(x)~A_i\gamma_j\gamma_5 ~\tfrac{1}{2}[\E^{\I \mathbf{Px}} 
 \overrightarrow{\nabla}_l  -\overleftarrow{\nabla}_l  \E^{\I \mathbf{Px}}] 
  u_s(x) -\{u_s\leftrightarrow d_s\}\quad(s=n,m,w)\FC\nonumber\\
{\cal O}_6^{s=n}(t)&=\tfrac{1}{\sqrt{2}}
[\pi^+(\mathbf{p_1})\pi^-(\mathbf{p_2})-\pi^-(\mathbf{p_1})\pi^+(\mathbf{p_2})]
\ ,\qquad \pi^{\pm}(\mathbf{p_i})=\sum_{\mathbf{x}} \bar q_{n}(x)  \gamma_5 
\tau^{\pm} \E^{\I \mathbf{ p_i x}} q_{n}(x)\FD
\end{align}
\end{widetext}
In the pion interpolator $\tau^\pm$ denote the corresponding
combination of Pauli matrices and the $\pi\pi$ interpolator ${\cal O}_6$
is always composed from narrow quarks. 
The covariant derivative (often denoted by $\overrightarrow{D}_i$)
\begin{equation}
\label{nabla}
\overrightarrow{\nabla}_i(\mathbf{x},\mathbf{y})=U_i(\mathbf{x},0)
\delta_{\mathbf{x}+\mathbf{i}, \mathbf{y}}-U_i^\dagger(\mathbf{x}-\mathbf{i},0)
\delta_{\mathbf{x}-\mathbf{i}, \mathbf{y}}\FC
\end{equation} 
is used in some  of the quark-antiquark interpolators (used already in a
number of lattice simulations, e.g. \cite{Gattringer:2008be,Engel:2010my}) and
will also be employed to prepare smeared quarks $q_s$ below.  It acts on the
spatial and color indices and leaves time and Dirac indices intact.  The linear
combinations in ${\cal O}_{4,5}$ are required for good $C$-parity.  The
polarization vector $\mathbf{A}$ of the quark-antiquark vector current  depends
on the total momentum $\mathbf{P}=\tfrac{2\pi}{L}\,\mathbf{d}$ as
\begin{align}
\label{polarizations}
\mathbf{d}&=(0,0,0)\ : \;\; \mathbf{A}=(0,0,1)\FC\ \mathbf{p_1}
=-\tfrac{2\pi}{L}\mathbf{A}\FC\ \mathbf{p_2}
=\tfrac{2\pi}{L}\mathbf{A}\FD\nonumber\\
\mathbf{d}&=(0,0,1)\ : \;\; \mathbf{A}
=\mathbf{d}\FC\qquad\ \  
 \mathbf{p_1}=\mathbf{0}\FC\ \ \qquad \mathbf{p_2}
 =\mathbf{P}\FD\nonumber\\
\mathbf{d}&=(1,1,0)\ : \;\; \mathbf{A}
=\mathbf{d}\FC\qquad\ \ 
\mathbf{p_1}=\mathbf{0}\FC\ \ \qquad \mathbf{p_2}
=\mathbf{P}\FD
\end{align}
Our choices for  $\pi\pi$ interpolators ${\cal O}_6$ \eq{interpolators} with
momentum projections for individual  pions   \eq{polarizations}  are  the
same as in \cite{Feng:2010es}:
\begin{itemize} 
\item For $\mathbf{d}=(0,0,0)$  with the symmetry group $O_h$ our
interpolator   transforms according to the three-dimensional representation $T_1^-$
(so just like $\mathbf{e_z}$)  under  elements of $O_h$. 
\item For $\mathbf{d}=(0,0,1)$  with the symmetry group $D_{4h}$ the
interpolator   transforms according to one-dimensional $A_2^-$ (like
$\mathbf{e_z}$)  under  elements of $D_{4h}$.
\item For $\mathbf{d}=(1,1,0)$  with the symmetry group $D_{2h}$ our
interpolator  transforms according to  one-dimensional $B_1^-$ (like
$\mathbf{e_x}+\mathbf{e_y}$)  under  elements of $D_{2h}$. Note that the 
interpolator $\mathcal{O}_6$
with $\mathbf{p_1}=(1,0,0)$ and $\mathbf{p_2}=(0,1,0)$ has the same total momentum, but it
has positive parity and it will not appear as an eigenstate for interpolators
with $B_1^-$ transformation properties.   
\end{itemize}

For the isovector pion $J^{PC}=0^{-+}$ correlation matrix  we use altogether
6 interpolators, using three smearing widths for each of the two Dirac structures,
\begin{eqnarray}\label{pion_interpolators}
{\cal O}^\pi_{type,s}(t)
=\sum _{\mathbf{x}}\bar u_s(x) \Gamma_{type}\E^{\I\mathbf{Px}} d_s(x)\FC\nonumber\\
\Gamma_1=\gamma_5,\quad \Gamma_2=\gamma_5\gamma_t,\quad s=n,\,m,\,w \FD
\end{eqnarray}

\subsection{Correlators and contractions }\label{sec:contractions}

\begin{figure}[bt]
\begin{center}
\includegraphics*[width=0.45\textwidth,clip]{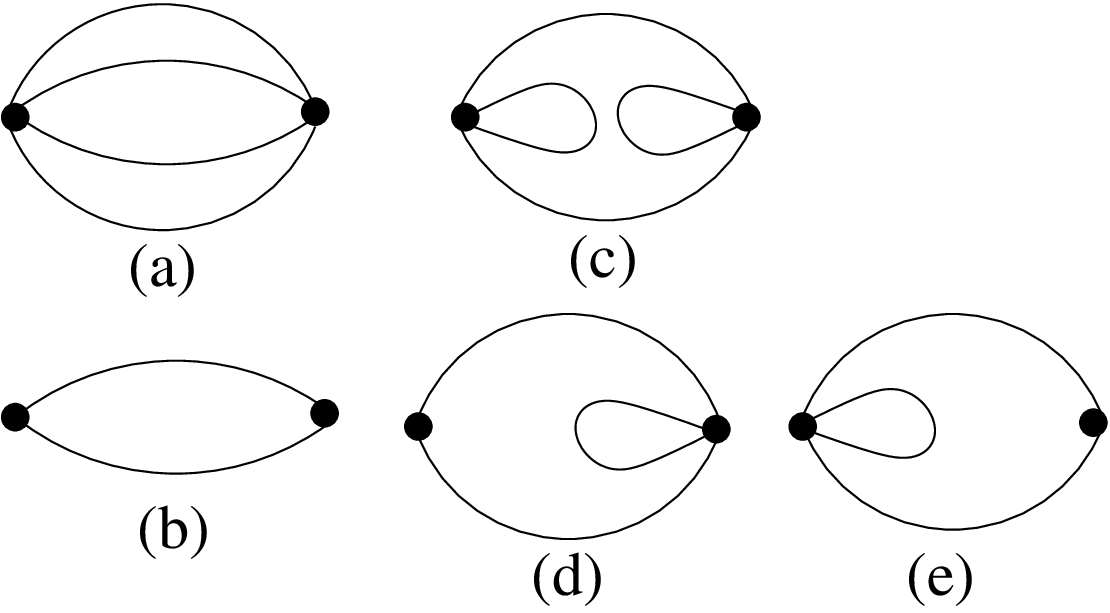}
\end{center}
\caption{ Contractions for our correlators with $\bar qq$ and $\pi\pi$
interpolators.}\label{fig:contractions}
\end{figure}

In the $\rho$ channel we compute $16\times 16$ correlation matrices for 
\begin{equation}
C_{jk}(t_f,t_i)=\langle 0|{\cal O}_j(t_f){\cal O}_k^\dagger (t_i)|0\rangle\
,\quad  j,\, k=1,\,16 \FC
\end{equation}
where the indices $j$ and $k$ stand for the combination $(type,s)$ in ${\cal
O}_{type,s}$ \eq{interpolators}.  These correlators involve 
(cf., Fig. \ref{fig:contractions}) connected contractions (a,b), singly
disconnected contractions (c), and contractions (d,e). Due to the 
momentum projections at the sink time slices $t_f$, the contractions (c) and (d)
in particular require the propagators  $M^{-1}$  from any spatial point at the
sink time slice $t_f=1,..,N_T$.

\subsection{Laplacian Heaviside smearing for quarks and the distillation
method}\label{sec:smearing}

Since calculating all elements of $M^{-1}$  for the fermion Dirac operator
matrix $M$ is
prohibitively time consuming, we apply the distillation method proposed in
\cite{Peardon:2009gh}. This method is based on a special kind of smearing for
quarks, that allows treatment of all necessary  contractions. All quarks are 
smeared according to a prescription similar to the conventional one
$q_{s}^{Gauss}(\mathbf{x},t)=\E^{\sigma_s \nabla ^2}q(\mathbf{x},t)$ where
$\nabla ^2$ denotes the 3D lattice Laplacian acting in a time slice.
The major simplification is due to  the 
spectral decomposition\footnote{$A$ is a
$N\times N$ matrix with eigenvalues $\lambda^{(k)}$ and eigenvectors $v^{(k)}$,
$Av^{(k)}=\lambda^{(k)}v^{(k)}$ ($k=1,..,N$).}  
\begin{equation}
f(A)=\sum_{k=1}^N f(\lambda^{(k)})~v^{(k)}v^{(k)\dagger}
\end{equation}  
for matrix $A\!=\!\nabla^2$ giving $\E^{\sigma_s \nabla^2 }\!=\!\sum_{1}^{N} \E^{\sigma_s \lambda^{(k)}}v^{(k)}v^{(k)\dagger}$. 
Here $\lambda^{(k)}$ and $v^{(k)}$ are
eigenvalues and eigenvectors of $\nabla^2$ \eq{nabla} which is a
$N_L^3N_c\times N_L^3N_c$ matrix  on a given gauge configuration 
\begin{equation}
\nabla^2_{\mathbf{x}c,\mathbf{x'}c'}(t) ~ v_{\mathbf{x'}c'}^{(k)}(t)
=\lambda^{(k)}(t)~ v_{\mathbf{x}c}^{(k)}(t)
\end{equation}
and all the resulting eigenvalues are negative. 
The
choice of smearing is arbitrary and  instead of this Gaussian smearing we use
the truncated spectral representation of the unit operator
(also called the Laplacian Heaviside (LapH) smearing), 
as proposed in \cite{Peardon:2009gh} and employed also in
\cite{Dudek:2009qf,Dudek:2010wm,Dudek:2010ew,Dudek:2011tt,Edwards:2011jj}
\begin{align}
\label{smearing}
&q_s\equiv \Theta(\sigma_s^2+\nabla^2)\;q=\sum_{k=1}^{N_cN_L^3} 
\Theta(\sigma_s^2+\lambda^{(k)}) ~v^{(k)} v^{(k)\dagger} \;q\FC\nonumber\\
 &q_s^{\alpha c}(\mathbf{x},t)=\sum_{k=1}^{N_v}v_{\mathbf{x}c}^{(k)}(t)~
 v_{\mathbf{x'}c'}^{(k)\dagger}(t)~q^{\alpha c'}(\mathbf{x'},t)\nonumber\\
 &\qquad\qquad\equiv 
\square_{\mathbf{xc,x'c'}}^{N_v} ~q^{\alpha c'}(\mathbf{x'},t)\FC\\
&\alpha,\alpha'=1,..,N_d\!=\!4\FC\quad c,c'=1,..,N_c\!=\!3~.\nonumber
\end{align}
The Heaviside smearing denoted by $\square^{N_v}$ is particularly suitable
since it cuts away the terms for $k>N_v$, where the number of eigenvectors $N_v$
kept in the sum depends on the chosen width $\sigma_{s=n,m,w}$. This choice of smearing reduces the number of needed inversions (per time slice,
Dirac index and configuration) from the prohibitively large number $N_L^3N_c$
(needed for the conventional all-to-all approach) to a manageable number $N_v\simeq
O(100)$.  

\begin{figure}[t]
\begin{center}
\includegraphics*[width=0.45\textwidth,clip]{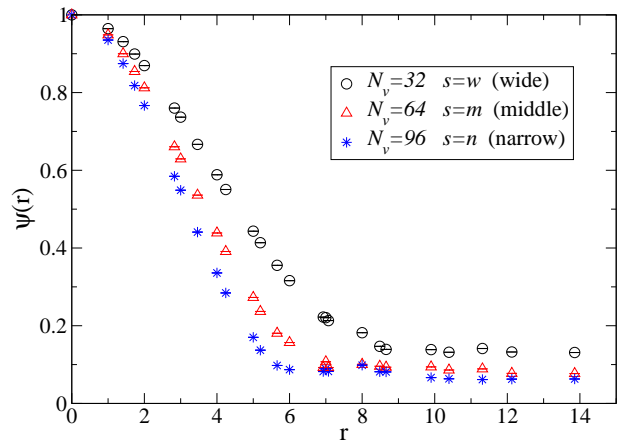}
\end{center}
\caption{The spatial distribution $\Psi(r)$ \eq{psi} of the distillation
operator $\square^{N_v}$ \eq{smearing} constructed from the eigenvectors corresponding to the $N_v$ lowest eigenvalues of the Laplace operator. The values are computed on each time slice of 49 configurations at distances along
the main axes and diagonals and plotted only until their respective symmetry
points. Circles (black), triangles (red) and stars (green) denote wide, middle
and narrow sources ($N_v=32, 64, 96$), respectively.}\label{fig:profiles}
\end{figure}

Different truncations correspond  to different effective smearing widths. 
We choose three smearing widths for quarks 
\begin{align}
&N_v=96\ \mathrm{for}\ s=n\ \mathrm{(narrow)}\FC\nonumber\\  
&N_v=64\ \mathrm{for}\ s=m\ \mathrm{(middle)}\FC\nonumber\\ 
&N_v=32\ \mathrm{for}\ s=w\ \mathrm{(wide)}\FC 
\end{align}
which lead to the spatial distributions \cite{Peardon:2009gh} of
\begin{equation}
\label{psi}
\Psi(r)=\sum_{\mathbf{x},t}\sqrt{\Tr_c[~\square_{\mathbf{x,x+r}}(t)~
\square_{\mathbf{x,x+r}}(t)~]}
\end{equation}
shown in Fig. \ref{fig:profiles}  

We build each interpolator \eq{interpolators} from quarks of the same width
for all three widths. 
This enlarges the variational basis and increases the possibility
for optimal eigensets.

\subsection{Evaluation of the correlators }\label{sec:eval_correl}

The interpolators  ${\cal O}_{1-5}$ given in \eq{interpolators} are linear combinations of
quark-antiquark  currents, which can be generally written as 
\begin{equation}
\label{current}
\bar Q_s^{\alpha 'c'}(\mathbf{x'},t)~~\Gamma_{\alpha'\alpha} 
~{\cal F}_{\mathbf{x'}\mathbf{x}}^{c'c}(t, \mathbf{p})~~
q_s^{\alpha c}(\mathbf{x},t)\FC\quad q,Q=u,d\FC 
\end{equation}
where the shape function ${\cal F}(t, \mathbf{p})$ incorporates the momentum
projection to $\mathbf{p}$  and the effect of covariant derivatives. Shape
functions ${\cal F}$ for our interpolators \eq{interpolators} are given in
\eq{app_F} of Appendix \ref{sec:app_contractions}.  The pion-pion
interpolator  ${\cal O}_6$ is a linear combination of products of two currents
\eq{current}.

After inserting the expression for smeared quarks $q_s$ of
\eq{smearing}  into interpolators \eq{interpolators},   all the
contractions for $C(t_f,t_i)$ can be expressed in terms of three
quantities $\Gamma$, $\phi$ and $\tau$, analogous to the original proposal
\cite{Peardon:2009gh} which considered only one smearing width. Correlators 
are expressed in terms of:  
\begin{itemize} 
\item Dirac matrices $\Gamma$ of size  $N_d\times N_d$.   \item The
interpolator shape matrices $\phi(t,{\cal F})$ are square matrices of size
$N_v\times N_v$ for an interpolator with a given smearing width $N_v$ 
\begin{equation}
\label{phi}
\phi^{k'k}(t,{\cal F})=\sum_{\mathbf{x'},\mathbf{x},c',c}~
v_{\mathbf{x'}c'}^{(k')\dagger}(t)~{\cal F}^{c'c}_{\mathbf{x'}\mathbf{x}}
(t,\mathbf{p})~v_{\mathbf{x}c}^{(k)}(t)~.
\end{equation}
Our $\phi$ is related to $\Phi$ in \cite{Peardon:2009gh} as $\Phi^{k'k}_{\alpha'\alpha}=\phi^{k'k} \Gamma_{\alpha'\alpha}$.
\item  The so-called perambulator matrices $\tau^{k'k}(t',t)$ 
\cite{Peardon:2009gh} denote the propagators from source of 
shape $v^k(t)$ to the sink of shape  $v^{k'}(t')$
\begin{equation}
\label{tau}
\tau_{\alpha'\alpha}^{k'k}(t',t)\equiv\!\!\!\! \sum_{\mathbf{x'},\mathbf{x},c',c}
v_{\mathbf{x'}c'}^{(k')\dagger}(t')~~(M^{-1})^{c'c}_{\alpha'\alpha}
(\mathbf{x'},t';\mathbf{x},t)~~v_{\mathbf{x}c}^{(k)}(t)~.
\end{equation}
Our  correlators depend on  the perambulators $\tau(t_f,t_i)$, $\tau(t_i,t_f)$,
$\tau(t_i,t_i)$, $\tau(t_f,t_f)$. These are  in general rectangular matrices
of sizes $N_dN_v^f\times N_dN_v^i,~ N_dN_v^i\times N_dN_v^f,~N_dN_v^i\times
N_dN_v^i$ and $N_dN_v^f\times N_dN_v^f$ respectively, where
$N_v^{i,f}=32,~64,~96$ denote the smearing widths of the source or sink.
\end{itemize}
The analytic expressions for the needed contractions (Fig.
\ref{fig:contractions}) in terms of $\Gamma$, $\tau$ and $\phi$ are given in
Appendix \ref{sec:app_contractions}. 

We precalculated and stored  the perambulators $\tau(t_f,t_i)$ from all source
times slices $t_i=1,..,N_T=32$ to all sink time slices $t_f=1,..,N_T=32$. This
allows us to compute all needed contractions for $C(t_f,t_i)$ straightforwardly. 
We sum\footnote{The sum plays the role of an average here.}   $C(t_f,t_i)$ over
all initial time slices $t_i$ to decrease the relative errors on the resulting
correlators $C(t=t_f-t_i)$.

We also sum over the results for the three $\rho$ polarizations
$\mathbf{A}=(0,0,1),~(0,1,0),~(1,0,0)$ for $\mathbf{d}=(0,0,0)$, or sum over the
directions $\mathbf{d}=(0,0,1),~(0,1,0),~(1,0,0)$ for $|\mathbf{d}|=1$, and  
over the directions $\mathbf{d}=(1,1,0),~(0,1,1),~(1,0,1)$ for
$|\mathbf{d}|=\sqrt{2}$. So, our final correlation matrices are 
\begin{equation}
C_{jk}(t=t_f-t_i)=\sum_{t_i=1,..,N_T} \sum_{\mathbf{A}\ \mathrm{or}\ \mathbf{d}} C_{jk}(t_f,t_i)\FD
\end{equation}
These correlation functions finally enter the variational analysis \eq{variational} 
to provide the energy levels. 

\subsection{Finite $N_T$ effects and the ``P+A'' trick}\label{subsec:PA}

Our  dynamical quarks have  antiperiodic boundary conditions  in
time. Using the valence quarks with the same antiperiodic boundary condition in time, we find
that the   finite time extent  $N_T=32$ ($T=3.96~$fm) severely affects the eigenvalues
$\lambda(t)$ near $t\simeq N_T/2=16$. There are two major sources for this: 
\begin{itemize}
\item The $\pi(\mathbf{p_1})\pi(\mathbf{p_2})$ state receives contributions from both pions
traveling forward or both traveling backward in time. But it also receives the contribution
from $\pi(\mathbf{p_1})$ traveling forward and $\pi(\mathbf{p_2})$ traveling backward in
time, and vice versa \cite{Prelovsek:2008rf,Detmold:2008yn}. As a result, the  cosh-type effective mass for 
some of the eigenvalues is not flat at $t>11$.
\item In the pion channel, the ground state starts to dominate the second largest eigenvalue (and vice
versa) at some moderate $t$ \cite{Gattringer:2008be,Gattringer:2008vj}.    
\end{itemize}

We use a previously applied  trick, which   effectively extends the time direction to $2N_T=64$ by
combining the periodic propagator $M^{-1}_P$ and antiperiodic propagator $M^{-1}_A$
(see for example \cite{Sasaki:2001nf,Detmold:2008yn}). All results in this paper have been obtained using the
so-called ``P+A'' propagators
\begin{align}
\label{P+A}
M^{-1}_{P+A}(t_f,t_i)=
\begin{cases}
\tfrac{1}{2}[M^{-1}_P(t_f,t_i) + M^{-1}_A(t_f,t_i)] & t_f\geq t_i\FC\\
\tfrac{1}{2}[M^{-1}_P(t_f,t_i) - M^{-1}_A(t_f,t_i)] & t_f< t_i \FD
\end{cases}
\end{align}
All our eigenvalues obtained from $M^{-1}_{P+A}$ agree with those obtained from $M^{-1}_{A}$
at $t\leq 11$. In the case of $M^{-1}_A$, the finite $T$ effects seriously affect some of the
eigenvalues for $t>11$. In the case of $M^{-1}_{P+A}$, the finite $T$ effects never show up in
any of the $\rho$ eigenvalues for $t\le 16$,  which allows us stable fit ranges at least until
$t=16$.  

The ``P+A trick'' is not a  valid field theoretic prescription, since the
valence quarks do not have the same periodicity as the dynamical quarks
(which remain antiperiodic in time). In practice, the pion
correlators with zero momentum, for example, are perfectly consistent with
periodicity $2N_T$, i.e. they are proportional to $\E^{-m_\pi t}+ \E^{-m_\pi
(2T-t)}$ and keep falling until $t=32$. We note that some of the nonzero
momentum $\rho$ correlators do not keep falling until  $t=32$, as would have
been expected in the case of the proper field theoretic  prescription. However, none
of the $\rho$ correlators shows  finite $T$ effects for $t<16$, which is the
time window used for our analysis.

\section{Computations}\label{Computations}

For the calculations presented here we use configurations generated for the
study of reweighting techniques in the p-regime of chiral perturbation theory. A
description of the normalized hypercubic smearing (nHYP smearing) used in the dynamic fermion action can be found
in \cite{Hasenfratz:2007rf}. Results from simulations with this action have
previously been published in \cite{Hasenfratz:2008ce,Hasenfratz:2008fg} and the
authors kindly provided the gauge configurations used in this study. The action
used to generate the gauge configurations containing $n_f=2$ flavors of
mass-degenerate light quarks is a tree-level improved Wilson-Clover action with
gauge links smeared using one level of nHYP smearing. Table \ref{gauge_configs}
lists the parameters used for the simulation along with the number of
(approximately independent) gauge configurations used and the pion mass
resulting from the determination of the lattice scale detailed in the next
subsection.

The gauge field obeys periodic boundary condition in time,  
while dynamical quarks are antiperiodic in time. As discussed in Sec. \ref{subsec:PA}, 
we compute and combine valence quark propagators with both antiperiodic and 
periodic boundary conditions.

\begin{table}[t]
\begin{ruledtabular}
\begin{tabular}{ccccccc}
$N_L^3\times N_T$ & $\kappa$ & $\beta$ & $a$[fm] & $L$[fm] & \#configs & $m_\pi$[MeV]\\ 
\hline
$16^3\times32$ & 0.1283 & 7.1 & 0.1239(13) & 1.98 & 280 & 266(3)(3) \\
\end{tabular}
\end{ruledtabular}
\caption{\label{gauge_configs}Configurations used for the current study. $N_L$
and $N_T$ denote the number of lattice points in spatial and time directions.
For the determination of the lattice spacing $a$ please refer to Sect. \ref{Computations}. The first error on $m_\pi$ is statistical while the second error is
from the determination of the lattice scale.}
\end{table}

On each gauge configuration we calculate the lowest 96 eigenvectors of the
lattice Laplacian on every time slice using a standard 3-point stencil.
Throughout, the gauge links are four dimensional nHYP smeared with the same parameters used
for generating the gauge configurations:
$(\alpha_1,\alpha_2,\alpha_3)=(0.75,0.6,0.3)$. For the calculation of the
eigenmodes and the interpolating fields containing covariant derivatives, we
also experimented with additional three-dimensional link-smearing (using regular HYP smearing)
and found only mild effects on the quality of simple meson two-point
correlators. We therefore opted to use no additional link-smearing. For the
calculation of the eigenmodes we use the PRIMME package
\cite{Stathopoulos:2009:PPI}. In particular the  routine 
\verb+JDQMR_ETol+
results in a
fast determination for a small to moderate number ($\mathcal{O}(10)$ to
$\mathcal{O}(100)$) of eigenmodes. For a larger number of eigenmodes the
Arnoldi/Lanczos method (and variants) eventually outperform this method. For the
methods implemented in PRIMME we also tried a preconditioner using Chebychev
polynomials, very similar to the method described in \cite{Morningstar:2011ka}.
While this greatly improved the performance of some methods, our preferred
method was largely unaffected and still outperformed all other PRIMME-methods
for a moderate number of eigenmodes.

For the determination of the quark propagators we use the \verb+dfl_sap_gcr+
algorithm provided in L\"uscher's DDHMC package
\cite{Luscher:2007se,Luscher:2007es}. Due to the large number of sources
necessary for the distillation approach, an inverter employing low-mode
deflation techniques is especially well suited. For the case presented here we
observed a speedup factor of approximately five compared to a BiCGStab 
algorithm without low-mode
deflation, while the computing time needed to generate the deflation subspace
was negligible compared to the actual calculation of quark propagators. Notice
that this difference gets more pronounced for the lighter quark masses needed
for future studies at or close to the physical point.

Statistical errors are determined with a single elimination jackknife procedure
throughout. When extracting energy levels we properly account for correlation in
Euclidean time $t$ by estimating the full covariance matrix in the given fit
interval. For the covariance matrix we use a jackknife estimate which is
calculated on the ensemble average 
only\footnote{This procedure has been referred
to as \emph{jackknife reuse} in \cite{Toussaint:2008ke}.}.

We determine the lattice spacing using the Sommer parameter
\cite{Sommer:1993ce}. We extract the static potential from planar Wilson loops
$W(r,t)$ obtained on gauge configurations smeared with hypercubic blocking
\cite{Hasenfratz:2001hp} with standard parameter values
$(\alpha_1,\alpha_2,\alpha_3)=(0.75,0.6,0.3)$. The potential is computed for
each value of $r$ from linear fits to $\log W(r,t)$ in the range $t=4\ldots 7$
and then  fitted to the lattice corrected form
\begin{equation}
V(r) = A + \frac{B}{r} + \sigma\,r + C\, \left(\left[\frac{1}{\mathbf r}\right]-
\frac{1}{r}\right)
\end{equation} 
in the range $1\le r\le 7$ or to the continuum form  (i.e., $C=0$) in the range
$2\le r\le 7$. Both values agree within less than one standard deviation. The
lattice corrections involves the lattice Coulomb potential $[1/\mathbf r]$
corrected for the hypercubic blocking
\cite{Hasenfratz:2001tw,Gattringer:2001jf}. To convert our numbers to physical
units (cf., Table \ref{gauge_configs}) we assume for the Sommer parameter the
value $r_0=0.48$ fm and obtain $a=0.1239(13)$ fm.

\section{Results}\label{Results}

\subsection{Pion results}

The pion energies are extracted from the variational analysis of the 6
interpolators given in \eq{pion_interpolators}.
The extracted pion mass and pion energies for the two lowest nonzero momenta
are given in Table \ref{tab:pi}, along with the analytic predictions from the
continuum and lattice dispersion relations.

\begin{table*}[t]
\begin{ruledtabular}
\begin{tabular}{c c c c  c c c c}
$\mathbf{P} \tfrac{L}{2\pi}$ & $t_0$ &interpol. & fit range &  $\chi^2$/d.o.f. & $E\,a$ (simul.)  & $E_{cont}^{d.r.}\,a$ & $E_{lat}^{d.r.}\,a$\vspace{3pt}\\
\hline
(0,0,0) & 3 & ${\cal O}_{1,2}^w{\cal O}_{1,2}^m{\cal O}_{1,2}^n$  & 8-14  & 1.57/5 & $m_\pi\,a=$0.1673(16)  & -- &-- \\
(0,0,1) & 3 & ${\cal O}_{2}^w{\cal O}_{2}^n$                                   & 12-17 & 0.98/4          & 0.4374(64) & 0.4268(65) & 0.4215(65) \\
(1,1,0) & 4 &  ${\cal O}_{2}^w{\cal O}_{1}^n$                                  & 8-13  & 1.31/4          & 0.5823(46) & 0.5800(48) & 0.5690(47) \\
\end{tabular}
\end{ruledtabular}
\caption{The ground state pion energy  extracted for three momenta:  $E$ is
extracted from the variational analysis using the chosen interpolator sets,
while $E^{d.r.}$ are obtained using the ground state pion mass and the
continuum and lattice dispersion
relations \eq{dispersion_lat}.}\label{tab:pi}
\end{table*}

\begin{figure}[bt]
\begin{center}
\includegraphics*[width=0.45\textwidth,clip]{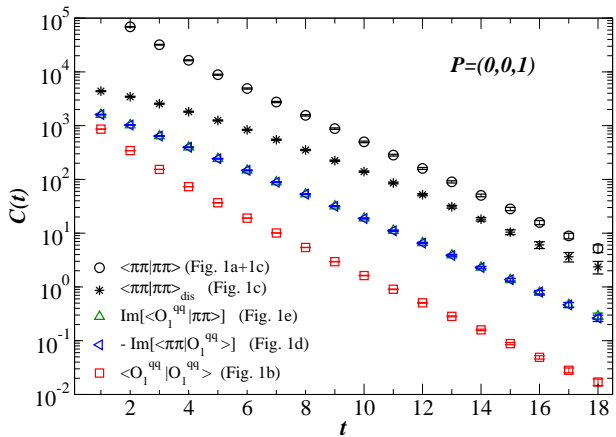}
\end{center}
\caption{An example of correlators for interpolators ${\cal O}_6=\pi\pi$ and ${\cal O}_{1}^n$ and their cross-correlators. }\label{fig:correlators}
\end{figure}

\begin{figure}[bt]
\begin{center}
\includegraphics*[width=0.48\textwidth,clip]{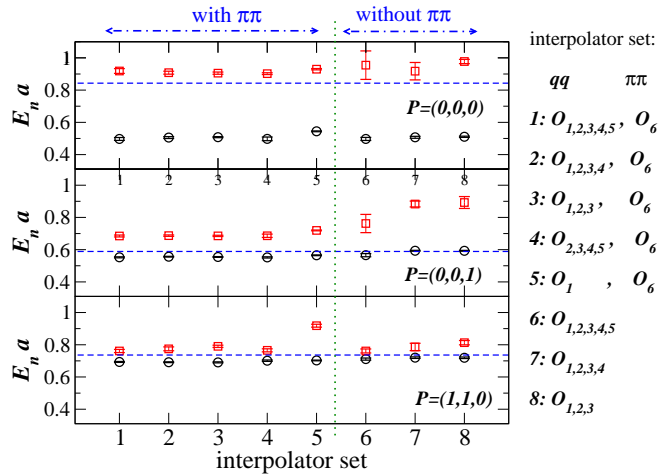}
\end{center}
\caption{The lowest two energy levels (circles denoting the ground state, 
squares the 1st excited state) extracted using different submatrices
(interpolators sets) of the full $16\times 16$ correlation matrix
(\ref{interpolators}), all for $t_0=4$. The horizontal dashed lines 
indicate the energy values for two 
noninteracting pions. The interpolators ${\cal O}_{1-5}$ have
$\bar q q$ valence structure, while ${\cal O}_6=\pi\pi$. All interpolators in
this plot are composed of narrow quarks $q_s=q_n$, with the exception of
interpolator set 3 which is ${\cal O}_1^w{\cal O}_2^m{\cal O}_3^n{\cal O}_6^n$.  
In order to make different interpolator choices comparable,  we use the same fit
range $t=7-10$ in the one-exponential correlated fit  for the purpose of this
figure (with exception of $E_2(d=0)$ obtained for $t=5-7$). }\label{fig:set_dep}
\end{figure}

\subsection{Rho meson results}

\subsubsection {Energy levels}

An example of the resulting correlators for interpolators $\pi\pi={\cal O}_6$ and
$\bar qq={\cal O}_1$, and their cross-correlators, are given in Fig.
\ref{fig:correlators}. 

Given our $16\times 16$ correlation matrices (\ref{interpolators}), we extracted
the two lowest energy levels for a number of different submatrices
(i.e., interpolator sets) of dimension $6\times 6$ or less.  Resulting levels for
eight different choices of interpolator sets are shown in Fig. \ref{fig:set_dep}. 
The extracted ground state energy is robust with regard to the choice of the interpolator
set, while the first excited energy is robust only if the interpolator set
includes the $\pi\pi$ interpolator and if the correlation matrix is larger than
$2\times 2$.  The first five choices include $\pi\pi$ in the interpolator basis,
while the last three do not.  The first excited energy for $d=(0,0,0)$ and
$d=(0,0,1)$ has much larger errors and is often substantially higher if $\pi\pi$
is not in the set. On the other hand, it seems that the  first excited energy in
the case $d=(1,1,0)$ can be extracted also without $\pi\pi$ interpolator in the
set.  The choice $set=5$ shows the result from the two-dimensional basis
$\pi\pi={\cal O}_6$ and $\bar qq={\cal O}_1$, which was used by some previous
simulations \cite{Aoki:2007rd,Aoki:2010hn,Feng:2010es,Frison:2010ws}. 
Figure \ref{fig:set_dep} indicates that
such a choice gives a reasonable estimate for the first excited energy in the
cases $d=(0,0,0)$ and $d=(0,0,1)$, while it gives a much higher energy for the
first excited state with $d=(1,1,0)$. Our study shows that a basis
larger than $2\times 2$ is needed to extract the  first excited level in this
case.

\begin{figure}[tp]
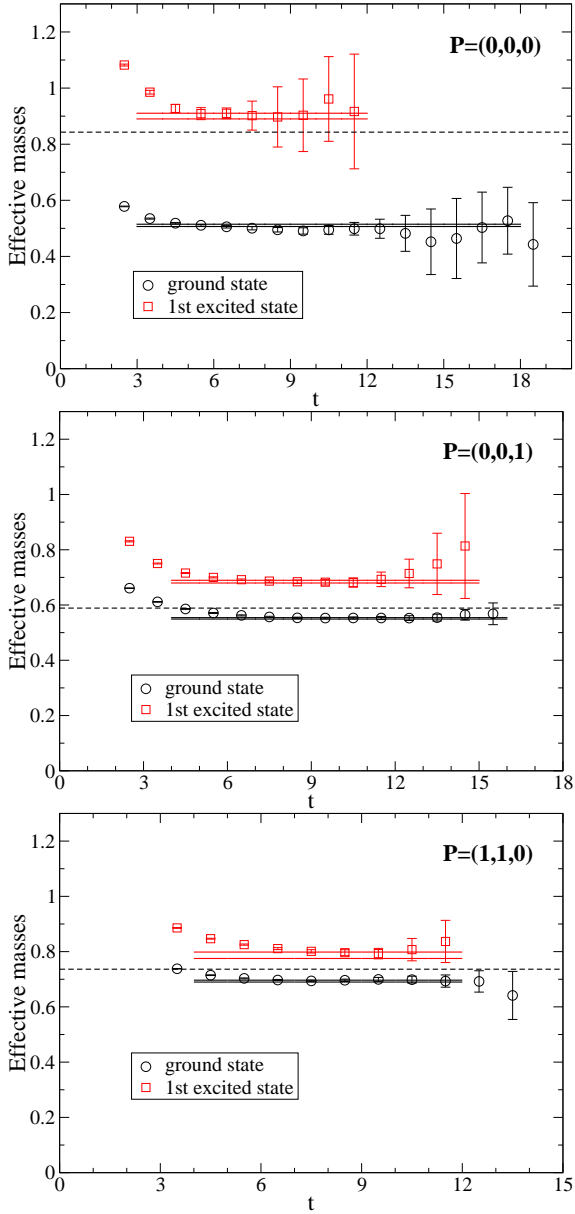

\begin{center}
\includegraphics*[width=0.42\textwidth,clip]{rho_0000000000111101_effmass.eps}
\includegraphics*[width=0.42\textwidth,clip]{rho_0000000000111101_effmass_001.eps}
\includegraphics*[width=0.42\textwidth,clip]{rho_0000000000111101_effmass_110.eps}
\end{center}
\caption{\label{fig:eff} The effective energies observed in the three momentum
frames (0,0,0), (0,0,1) and (1,1,0), based on diagonalization of a correlation matrix
with 4 or 5 operators, listed in Table \ref{tab:results}. The horizontal bands indicate the resulting energy levels
derived from two-exponential fits to $\lambda_i(t)$ as discussed in the text. The dashed lines
indicate the noninteracting two-pion levels as determined from the energies
$E_{lat}^{d.r.}a$ in Table \ref{tab:pi}.}
\end{figure}

Given that our lowest two energy levels are robust with respect to the choice of
interpolator set provided the set is large enough and contains the $\pi\pi$
interpolator, we present the final interpolator set choices in Table
\ref{tab:results}. The corresponding  effective masses  for our preferred
interpolator choices are shown in Fig. \ref{fig:eff}.  The final values for the
six energy levels in Table \ref{tab:results}  are extracted using correlated
two-exponential fits with $t_0$ as indicated in the table and starting at a rather small time separation $t$. We verified that
the extracted levels agree with  results obtained from one-exponential fits
starting at larger $t$ and using $t_0=[2,5]$. 

\begin{table*}[t]
\begin{ruledtabular}
\begin{tabular}{rccrclllll}
$\mathbf{P}\frac{L}{2\pi}$ & level $n$ & $t_0$ &interpol. & fit range & $E_na$ & $\chi^2$/d.o.f. 
& $a\,p^*$ & $s\,a^2$ & $\delta$\vspace{3pt}\\
\hline
(0,0,0) &1& 2 & ${\cal O}^n_{1,2,3,4,6}$&3-18&0.5107(40)&6.10/12&0.1940(29) & 0.2608(41) & 130.56(1.37)\\
(0,0,0) &2& 2 & ${\cal O}^n_{1,2,3,4,6}$&3-12&0.9002(101)&0.85/6  & 0.4251(58)& 0.8103(182) & 146.03 (6.58) [*]\\
(0,0,1) &1& 2 & ${\cal O}^n_{1,2,3,4,6}$&4-16&0.5517(26)  &4.06/9  & 0.1076(36)& 0.1579(29) &  3.06 (0.06)\\
(0,0,1) &2& 2 & ${\cal O}^n_{1,2,3,4,6}$&4-15&0.6845(49)  &3.10/8  & 0.2329 (40)& 0.3260(69) &156.41(1.56)\\
(1,1,0) &1& 3 & ${\cal O}^n_{1,2,3,6}$&4-12&0.6933(33)  &4.33/5  & 0.1426(42)& 0.1926(49) & 6.87(0.38)\\
(1,1,0) &2& 3 & ${\cal O}^n_{1,2,3,6}$&4-12	&0.7868(116)&2.34/5 & 0.2392(101) & 0.3375(191) & 164.25(3.53)
\end{tabular}
\end{ruledtabular}
\caption{\label{tab:results}
Final results for the lowest two $\rho$ energy levels, all obtained using 2-exp
correlated fits with given $\chi^2$/d.o.f.. The choice of interpolator basis
\eq{interpolators} is indicated. The pion momenta $a\,p^*$ in the CMF and
scattering phases $\delta$ are obtained using the lattice dispersion relation
\eq{dispersion_lat} and $m_\pi\,a$ in Table \ref{tab:pi}. 
 The state $E_2(P=0)$ is above the $4\pi$ threshold and is denoted by a star.}
\end{table*}

\begin{figure}[tp]
\begin{center}
\includegraphics*[width=0.45\textwidth,clip]{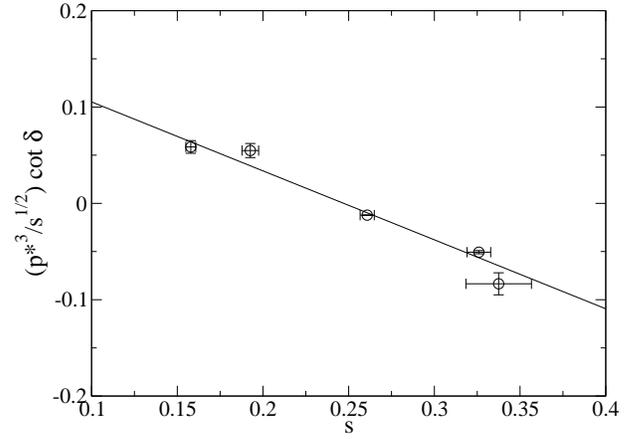}
\end{center}
\caption{\label{fig:cot_delta}
Our data for $((a\,p^*)^3/\sqrt{s\,a^2})\,\cot \delta(s)$  as a function of
$s\,a^2$, fitted to straight line  behavior according  to \eq{delta_fit}. The
fit has $\chi^2/$d.o.f.=7.42/3 and gives $g_{\rho\pi\pi}=5.13(20)$ and $m_\rho\,
a=0.4972(42)$. The states $E_n(\mathbf{d})$ corresponding to various points  can
be deduced by the value of $s$ in Table \ref{tab:results}. 
The plot data is shown in units of the lattice spacing.
}
\end{figure}

\begin{figure}[tp]
\begin{center}
\includegraphics*[width=0.45\textwidth,clip]{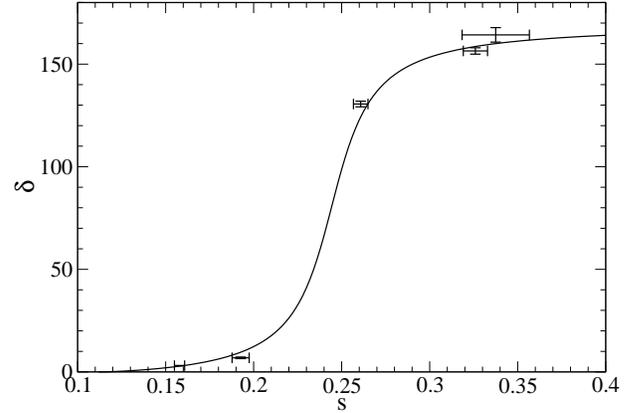}
\end{center}
\caption{\label{fig:phase_shift}
The p-wave phase-shift values compared with the result from the fit to 
\eq{delta_fit} in Fig. \ref{fig:cot_delta} for $g_{\rho\pi\pi}=5.13$ and
$m_\rho\,a=0.4972$. The states $E_n(\mathbf{d})$ corresponding to various points 
can be deduced by the value of $s$ in Table \ref{tab:results}. }
\end{figure}

\subsubsection{Phase-shifts and resonance parameters}

Each of the energy levels of Table \ref{tab:results} gives the value of the scattering phase
shift $\delta(s)$ at a different pion CMF momentum $p^*$. We employed the lattice
dispersion relation \eq{dispersion_lat} to get $p^*=\tfrac{2\pi}{L}q$ and
used the phase-shift formulas in Sect. \ref{sec:tools_phase_shifts} to get
$\delta(q^2)$. Our results, including jack-knife error estimates, are also given in
Table \ref{tab:results}. 

The resulting phase-shift is related to the  relativistic Breit-Wigner form for the elastic p-wave amplitude 
in the resonance region \cite{Nakamura:2010zzi} 
\begin{equation}
\label{amplitude}
a_1=\frac{-\sqrt{s}\,\Gamma(s)}{s-m_\rho^2+\I \sqrt{s}\,\Gamma(s)}=
\E^{\I\delta(s)}\sin \delta(s)\FC
\end{equation}
where $s=E_{CM}^2$ is the Mandelstam variable and $m_\rho^2$ is the resonance
position. Relation \eq{amplitude} can be conveniently written for later use as
\begin{equation}
\label{amplitude1}
\sqrt{s}\,\Gamma(s)\,\cot \delta(s)=m_\rho^2-s\FC
\end{equation}
and the  decay width $\Gamma(s)$ is expressed in terms of  the coupling constant $g_{\rho\pi\pi}$, taking into account the $\pi\pi$ phase space  \cite{Brown:1968zz,Renard:1974cz} 
\begin{equation}
\label{width}
\Gamma(s)=\frac{{p^*}^3}{s} \frac{g_{\rho\pi\pi}^2}{6\pi}\FC
\end{equation}
where the $\rho$ width $\Gamma_\rho=\Gamma(m_\rho^2)$ is evaluated at the
resonance position. 

The final relation, the so-called effective range formula,  combines
(\ref{amplitude1},\ref{width}) and is valid in the elastic region 
$s< (4 m_\pi)^2$,
\begin{equation}\label{delta_fit}
\frac{{p^*}^3}{\sqrt{s}}\,\cot \delta(s)= \frac{6\pi}{g_{\rho\pi\pi}^2} ( m_\rho^2 - s )~.
\end{equation}
It allows a linear fit for the two unknown parameters  $6\pi/g_{\rho\pi\pi}^2$
and  $6\pi\, m_\rho^2/g_{\rho\pi\pi}^2$.  Values of $s$, $p^*$ and $\delta$ for
the energy levels $E_n$ are given in Table \ref{tab:results} and appropriate
combinations (\ref{delta_fit}) are plotted  in Fig. \ref{fig:cot_delta}. In the
fit and in the figures we do not include the first excited state with $P=0$,
since this lies above the $4 \pi$ inelastic threshold.

Figure \ref{fig:cot_delta} shows the result of the linear fit to the data, 
giving our final result for $g_{\rho\pi\pi}$ and the mass of the $\rho$ resonance 
(at our $m_\pi=266(3)(3)\;$MeV),
\begin{align}
\label{g}
g_{\rho\pi\pi}&=5.13(20)\FC\\
m_\rho\,a&=0.4972(42)\FC \qquad m_\rho=792(7)(8)\;\mathrm{MeV}\FD\nonumber
\end{align}
Figure \ref{fig:phase_shift} exhibits the corresponding phase-shift in the
resonance region.  
The values (\ref{g}) are obtained using the lattice
dispersions relation (\ref{dispersion_lat}). Given the systematic
uncertainty with simulations on a single ensemble, they agree reasonably well
with the results $g_{\rho\pi\pi}=5.60(18)$ and $m_\rho\,a =0.4833(41)$
obtained using the naive dispersion relation.

The value of the coupling (\ref{g}) is near the experimental value
$g_{\rho\pi\pi}^{exp}\approx 5.97$. Our coupling is also compatible with the results in \cite{Aoki:2007rd,Aoki:2010hn} within the errors given there. Note that \cite{Aoki:2007rd,Aoki:2010hn} computed the coupling at a larger pion mass. In \cite{Feng:2010es} a larger value $g_{\rho\pi\pi}=6.77(67)$ and a substantially
larger $m_\rho=980$ MeV are observed at similar pion mass $m_\pi=290$ MeV. Our $m_\rho$ is close to the prediction of the unitarized one-loop\footnote{The two loop result strongly depends on a number of poorly known Low Energy Constants, which are fixed in \cite{PelaezRios2010} also by using the lattice data on $m_\rho$, so the comparison to the two-loop result is not appropriate.} Chiral  Perturbation theory (ChPT), which leads to about $m_\rho\simeq 800 \mathrm{MeV}$ at $m_\pi\simeq 266 \mathrm{MeV}$ \cite{Hanhart2008,PelaezRios2010}. We also compared our $\delta(s)$ with the prediction of unitarized ChPT, recalculated  for our $m_\pi=266 \mathrm{MeV}$ by the authors of \cite{Nebreda2011}: we find good agreement for $\sqrt{s}<m_\rho$ and reasonable agreement with one-loop results for $\sqrt{s}>m_\rho$.

Since the width is crucially influenced by the $\pi\pi$-phase space, this number
derived for our pion mass comes out significantly smaller than the experimental
value, so we present only $g_{\rho\pi\pi}$. This dimensionless coupling is
expected  to be almost independent of $m_\pi$ \cite{PelaezRios2010}, which was also explicitly verified
in a study for several pion masses \cite{Feng:2010es}.

\section{Conclusions and outlook}

Extracting scattering phase-shifts and resonance properties is one of the most
challenging problems in hadron spectroscopy based on lattice QCD.  We combine
several sophisticated tools  to approach this problem:  L\"uscher's phase-shift
relations for finite-volume lattices, moving frames and variational analysis of
correlation matrices, where a number of quark-antiquark and $\pi\pi$
interpolators with quantum numbers $I(J^{PC})=1(1^{--})$ are used.  All needed
contractions are evaluated using the distillation method with the Laplacian
Heaviside smearing of quarks. We find that these tools lead to precise values of
the p-wave phase-shift for $\pi\pi$ scattering at  five values of pion relative
momenta in the vicinity of the resonance. This allows a determination of the
$\rho$ resonance parameters $m_\rho$ and $\Gamma_\rho$ at our value of $m_\pi$. 

The simulation is performed on an ensemble 
\cite{Hasenfratz:2008ce,Hasenfratz:2008fg} of 280 gauge configurations with two
mass-degenerate dynamical clover-improved Wilson fermions. The pion mass $m_\pi$
is roughly 266 MeV, the lattice volume $V$ is $16^3\times 32$ and the spatial
extent of the lattice is $L\simeq 1.98~$fm. The exponentially suppressed
finite-volume corrections may not be completely negligible at our $m_\pi L\simeq
2.68$ and future simulations will have to improve on this.  Larger lattices will
necessitate stochastic estimation techniques to avoid the unsatisfactory scaling
of full distillation with the lattice volume. Such a method has recently been
provided in \cite{Morningstar:2011ka}. In the present study we calculated the
quark propagation by calculating the distillation perambulators on all time
slices, which is not very economical and only feasible in small volumes.

Along the way, we explore how well the lowest two energy levels can be
obtained without the $\pi\pi$ interpolators  in the variational basis. We
also propose how to treat interpolators of different smearing widths in the
same variational basis within the distillation method.

We demonstrate  that a relatively accurate determination of the resonance
parameters is possible with present day techniques, within the limitation
of small $m_\pi L$.  For our pion mass we obtain the  resonance mass
$m_\rho=792(7)(8)\;$MeV and the $\rho\to\pi\pi$ coupling 
$g_{\rho\pi\pi}=5.13(20)$, which is close to the experimental value
$g_{\rho\pi\pi}^{exp}\approx 5.97$. We prefer to give the coupling, since
the actual width $\Gamma_\rho$ is strongly  affected by the phase space,
which is small due to the large value of our pion mass.

Following the pion, the rho is the most prominent meson. With sharpened
tools it is now becoming possible to analyze its decay properties. The
present study of the $\rho$ resonance gives us confidence that  similar
techniques can be applied to also extract the resonance parameters of  some
other hadronic resonances and we intend to pursue research along these
lines in the near future.

\acknowledgments
First of all, we would like to kindly thank Anna Hasenfratz for providing
the gauge configurations used for this work.  We would like to thank Gilberto Colangelo, Georg
Engel, Xu Feng, Christof Gattringer, Jose Pelaez, Akaki Rusetski, Igor Sega, Gerrit
Schierholz and Richard Woloshyn for valuable discussions. The calculations
have been performed on the theory cluster at TRIUMF and on local clusters at
the University of Graz and Ljubljana.  We thank these institutions for
providing support. This work is supported by the Slovenian Research Agency,
by the European RTN network FLAVIAnet (contract number MRTN-CT-035482), by
the Slovenian-Austrian bilateral project (contract number  BI-AT/09-10-012)
and by the Natural Sciences and Engineering Research Council of Canada
(NSERC).

\begin{appendix}

\section{Contractions in the distillation method}
\label{sec:app_contractions}

Here we provide the analytic expressions for correlators $C(t_f,t_i)=\langle
{\cal O}_{f}(t_f)~{\cal O}_{i}(t_i)\rangle$ that follow from general quark
antiquark interpolators with $|I,I_3\rangle=|1,0\rangle$ (with examples given
by ${\cal O}_{1-5}$ in \eq{interpolators})
\begin{align}
\label{app_qq}
{\cal O}^{\bar qq}_f(t_f)&=\frac{1}{\sqrt{2}}\bigl[\bar u_{s_f}(t_f)~
\Gamma_f^0 ~{\cal F}^0_f(t_f,\mathbf{P}) ~u_{s_f}(t_f)-
\{u\leftrightarrow d\}\bigr]\FC\nonumber\\
\quad {\cal O}^{\bar qq}_i(t_i)&=\frac{1}{\sqrt{2}}\bigl[\bar u_{s_i}(t_i)~
\Gamma^0_i ~{\cal F}^0_i(t_i,-\mathbf{P}) ~u_{s_i}(t_i)-
\{u\leftrightarrow d\}\bigr]\FC
\end{align} 
and general meson-meson ($MM$) interpolators with  
$|I,I_3\rangle=|1,0\rangle$ (with example given by ${\cal O}_{6}$ in 
\eq{interpolators})
\begin{align}
\label{app_MM}
   {\cal O}^{MM}_{f}(t_f)&=\frac{1}{\sqrt{2}}\bigl[\bar d_{s_f}(t_f)~
   \Gamma_{1f} ~{\cal F}_f(t_f,\mathbf{p_{1f}})~u_{s_f}(t_f)\times\nonumber\\
   &\bar u_{s_f}(t_f)~\Gamma_{2f}~{\cal F}_f(t_f,\mathbf{p_{2f}})~
   d_{s_f}(t)-\{u\leftrightarrow d\}\bigr]\FC\nonumber\\
{\cal O}^{MM}_{i}(t_i)&=\frac{1}{\sqrt{2}}\bigl[\bar u_{s_i}(t_i)~
\Gamma_{1i} ~{\cal F}_i(t_i,-\mathbf{p_{1i}})~d_{s_i}(t_i)\times\nonumber\\
&\bar d_{s_i}(t_i)~\Gamma_{2i}~{\cal F}_i(t_i,-\mathbf{p_{2i}})~
u_{s_i}(t)-\{u\leftrightarrow d\}\bigr]\FD
  \end{align}
The subscripts $s_i,~s_f=n,m,w$ denote the smearing width of the sink and
source. The superscript ``0'' denotes that $\Gamma^0$ and ${\cal F}^0$ apply to
$\bar qq$ interpolators, while $\Gamma$ and ${\cal F}$ without superscript
apply to meson-meson interpolators.  

The shape functions ${\cal F}$ \eq{current} for our interpolators
\eq{interpolators} are of three types
\begin{align}
\label{app_F}
\mathrm{no}\ \nabla:
{\cal F}^{c'c}_{\mathbf{x'x}}(t,\mathbf{p})&=
 \delta_{c'c}\delta_{\mathbf{x'x}}
\E^{\I \mathbf{px}} \FC\nonumber\\
\mathrm {for}\ \mathbf{\nabla}:
{\cal F}^{c'c}_{\mathbf{x' x}}(t,\mathbf{p})&=
\tfrac{1}{2}\bigl[
\E^{\I \mathbf{px}}(\overrightarrow{\nabla}_j)^{c'c}_{\mathbf{x'x}}(t)-
(\overleftarrow{\nabla}_j)^{c'c}_{\mathbf{x'x}}(t)
\E^{\I \mathbf{px}}\bigr]\FC\nonumber\\
\mathrm{for}\ \mathbf{\nabla\nabla}:
{\cal F}^{c'c}_{\mathbf{x'x}}(t,\mathbf{p})&=\sum_{j=1,2,3} 
(\overleftarrow{\nabla}_j)^{c'c_0}_{\mathbf{x'x_0}}(t)
\E^{\I \mathbf{px_0}}(\overrightarrow{\nabla}_j)^{c_0c}_{\mathbf{x_0x}}(t)
\FC
\end{align}
and we use the first choice (without $\nabla$) within our meson-meson
interpolators. 

The contractions in Fig. \ref{fig:contractions} are expressed in terms of the
perambulators $\tau$ \eq{tau}, interpolator shape matrices $\phi$ \eq{phi}
and Dirac matrices $\Gamma$, which are presented in Section
\ref{sec:eval_correl} of the main text. The analytic expressions for the
contractions are 
\begin{widetext}
\begin{align}
&\langle {\cal O}^{\bar qq}_{f}(t_f) 
{\cal O}^{\bar qq}_{i}(t_i)\rangle=C^{Fig.1b}(t_f,t_i)
=-\Tr \bigl[~\tau(t_i,t_f)~\Gamma_f^0~\phi(t_f,{\cal F}_f^0(\mathbf{P}))~
\tau(t_f,t_i)~\Gamma_i^0~\phi(t_i,{\cal F}_i^0(-\mathbf{P}))~\bigr]
\FD
\end{align}
\begin{align}
&\langle {\cal O}^{MM}_{f}(t_f) 
{\cal O}^{\bar qq}_{i}(t_i)\rangle=C^{Fig.1d}(t_f,t_i)=\nonumber\\
&\Tr\bigr[~\tau(t_i,t_f)~\Gamma_{1f}~\phi(t_f,{\cal F}(\mathbf{p_{1f}}))~
\tau(t_f,t_f)~\Gamma_{2f}~\phi(t_f,{\cal F}(\mathbf{p_{2f}}))~
\tau(t_f,t_i)~\Gamma_i^0~\phi(t_i,{\cal F}_i^0(-\mathbf{P}))~\bigr]
\nonumber\\
&\qquad  ~-~\{ \mathbf{p_{1f}}\leftrightarrow \mathbf{p_{2f}}\FC\ 
\Gamma_{1f}\leftrightarrow \Gamma_{2f}\}
\FD
\end{align}
\begin{align}
&\langle {\cal O}^{\bar qq}_{f}(t_f)
 {\cal O}^{MM}_{i}(t_i)\rangle=C^{Fig.1e}(t_f,t_i)=\nonumber\\
&-\Tr\bigr[~\tau(t_f,t_i)~\Gamma_{1i}~\phi(t_i,{\cal F}(-\mathbf{p_{1i}}))~
\tau(t_i,t_i)~\Gamma_{2i}~\phi(t_i,{\cal F}(-\mathbf{p_{2i}}))~\tau(t_i,t_f)
~\Gamma_f^0~\phi(t_f,{\cal F}_f^0(\mathbf{P}))~\bigr]\nonumber\\
&\qquad  ~+~\{ \mathbf{p_{1i}}\leftrightarrow \mathbf{p_{2i}}\FC\ 
\Gamma_{1i}\leftrightarrow \Gamma_{2i}\}
\FD
\end{align}
\begin{align}
&\langle {\cal O}^{MM}_{f}(t_f) 
{\cal O}^{MM}_{i}(t_i)\rangle=C_{con}^{Fig.1a}(t_f,t_i)+
C_{dis}^{Fig.1c}(t_f,t_i)\FD\nonumber\\
&~\nonumber\\
&C_{con}^{Fig.1a}(t_f,t_i)=\Tr\bigl[~\tau(t_i,t_f)~\Gamma_{1f}~
\phi(t_f,{\cal F}_f(\mathbf{p_{1f}}))~\tau(t_f,t_i)~\Gamma_{1i}~
\phi(t_i,{\cal F}_i(-\mathbf{p_{1i}}))~\bigr]\nonumber\\
&\qquad \quad \quad\qquad \times ~  \Tr\bigl[~\tau(t_i,t_f)~\Gamma_{2f}~
\phi(t_f,{\cal F}_f(\mathbf{p_{2f}}))~\tau(t_f,t_i)~\Gamma_{2i}~
\phi(t_i,{\cal F}_i(-\mathbf{p_{2i}}))~\bigr]\nonumber\\
&\qquad \quad \qquad\qquad \  ~-~\{ \mathbf{p_{1i}}\leftrightarrow 
\mathbf{p_{2i}}\FC\ \Gamma_{1i}\leftrightarrow \Gamma_{2i}\}\FC\nonumber\\
&C_{dis}^{Fig.1c}(t_f,t_i)=\Tr\bigl[~\tau(t_i,t_f)~\Gamma_{2f}~
\phi(t_f,{\cal F}_f(\mathbf{p_{2f}}))~\tau(t_f,t_f)~\Gamma_{1f}~
\phi(t_f,{\cal F}_f(\mathbf{p_{1f}}))~\nonumber\\
&\qquad \qquad \qquad\qquad \ \cdot ~\tau(t_f,t_i)~\Gamma_{2i}~
\phi(t_i,{\cal F}_i(-\mathbf{p_{2i}}))~\tau(t_i,t_i)~\Gamma_{1i}~
\phi(t_i,{\cal F}_i(-\mathbf{p_{1i}}))~\bigr]\nonumber\\
& \qquad \qquad\qquad + ~\Tr\bigl[~\tau(t_i,t_f)~\Gamma_{1f}~
\phi(t_f,{\cal F}_f(\mathbf{p_{1f}}))~\tau(t_f,t_f)~\Gamma_{2f}~
\phi(t_f,{\cal F}_f(\mathbf{p_{2f}}))~\nonumber\\
&\qquad \qquad \qquad\qquad \quad \cdot ~\tau(t_f,t_i)~\Gamma_{1i}~
\phi(t_i,{\cal F}_i(-\mathbf{p_{1i}}))~\tau(t_i,t_i)~\Gamma_{2i}~
\phi(t_i,{\cal F}_i(-\mathbf{p_{2i}}))~\bigr]\nonumber\\
&\qquad \quad \qquad\qquad\qquad \  ~-~\{ \mathbf{p_{1i}}\leftrightarrow 
\mathbf{p_{2i}}\FC\ \Gamma_{1i}\leftrightarrow \Gamma_{2i}\}\FD
\end{align}
\end{widetext}
and can be generally used for the interpolators of the form \eq{app_qq} or
\eq{app_MM}, or their cross-correlators. 
 
\end{appendix}
\clearpage


%

\clearpage
\newpage

\begin{center}
{\Large{\textbf{Erratum: Coupled channel analysis of the $\rho$ meson decay in lattice QCD}}}
\end{center}

\begin{figure}[bp]
\begin{center}
\includegraphics*[width=0.48\textwidth,clip]{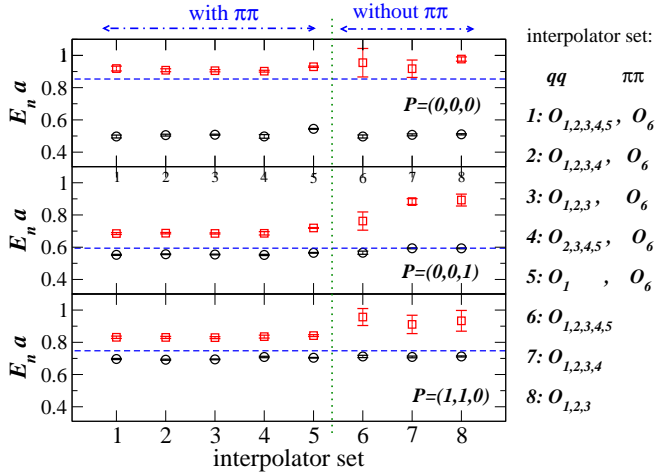}
\end{center}
\caption{The lowest two energy levels (circles denoting the ground state, 
squares the 1st excited state) extracted using different sub-matrices
(interpolators sets) of the full $16\times 16$ correlation matrix. The dashed lines have been changed to show values using the continuum dispersion relation.}\label{fig:set_dep}
\end{figure}

\begin{figure}[bhp]
\begin{center}
\includegraphics*[width=0.45\textwidth,clip]{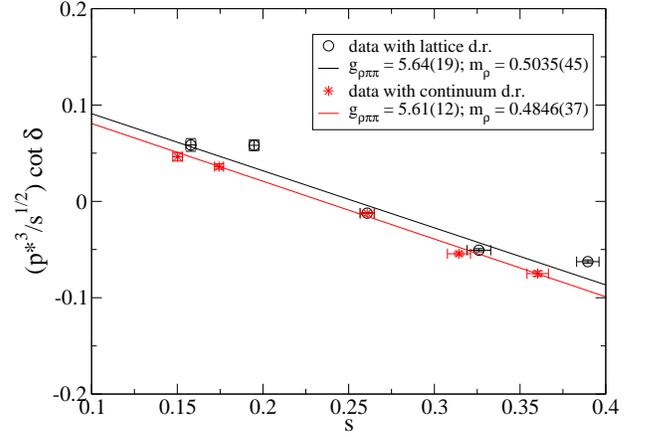}
\end{center}
\caption{Data for $((a\,p^*)^3/\sqrt{s\,a^2})\,\cot \delta(s)$  as a function of
$s\,a^2$, fitted to straight line behavior. We also include data using the lattice dispersion relation to allow for a comparison with the previously published results.}\label{fig:kmatrix}
\end{figure}

\begin{table*}[htb]
\begin{ruledtabular}
\begin{tabular}{rccrclllll}
$\mathbf{P}\frac{L}{2\pi}$ & level $n$ & $t_0$ &interpol. & fit range & $E_na$ & $\chi^2$/d.o.f. 
& $a\,p^*$ & $s\,a^2$ & $\delta$\vspace{3pt}\\
\hline
(1,1,0) &1& 2 & ${\cal O}^n_{1,2,3,4,6}$&3-13&0.6948(19)  &4.20/7  & 0.1249(29)& 0.1743(27) & 7.37(0.11)\\
(1,1,0) &2& 2 & ${\cal O}^n_{1,2,3,4,6}$&3-11&0.8177(38)  &1.97/5  & 0.2492(32)& 0.3603(63)& 161.03(1.20)
\end{tabular}
\end{ruledtabular}
\caption{\label{tab:results}
Results for the lowest two $\rho$ energy levels, all obtained using the continuum dispersion relation and 2-exp
correlated fits with given $\chi^2$/d.o.f.. For further explanation please refer to the published paper.}
\end{table*}

In our computer code used for generating the published data, we incorrectly assumed that $\langle \bar q \gamma_x q|\bar q \gamma_y q\rangle\!=\! 0$  (and analogous for other types of $\bar qq$ interpolators) -- which is true for irrep $T_1^-$ with $\mathbf{P}\!=\!(0,0,0)$ -- is also true for irrep $B_1^-$ with $\mathbf{P}\!=\!\tfrac{2\pi}{L}(1,1,0)$. This error  mildly influences only results for $\mathbf{d}\!=\!\tfrac{L}{2\pi}\mathbf{P}\!=\!(1,1,0)$. This modifies our preferable choice of dispersion relation: we regarded the lattice dispersion relation derived from  nearest neighbor central difference (Eqs. (10,11) in the paper) as more suitable, while the corrected data suggests that the continuum dispersion relation (Eqs. (6,9)) is more suitable. 

Figure \ref{fig:set_dep} shows an update for Fig. 4. of the paper; the corrected data is labeled by $\mathbf{P}\!=\!(1,1,0)$. While the ground state is only affected within the statistical uncertainty (the difference is not significant even when taking into account correlations), the excited state is affected strongly (Figure 5.c., which we omit for brevity, is affected similarly). We previously stated that interpolator set 5 ($O_1$, $O_6$) leads to a much higher energy for the first excited state in the frame with $\mathbf{d}\!=\!(1,1,0)$. With the corrected data, this basis still suffers from strong excited state contaminations but the results are much closer to those with a larger basis. Further conclusions drawn from Fig. 4 in the paper remain unchanged.

Table \ref{tab:results} shows the corrected data (using the continuum dispersion relation) and the values supersede the corresponding entries in Table III of the paper.

Figure \ref{fig:kmatrix} shows a correction to Fig.  6 of the paper. In addition to the results based on  lattice dispersion relation (black circles), we also show the continuum dispersion relation (red stars). Whereas the quality of the fit as expressed by the $\chi^2$/d.o.f. is quite bad using the lattice dispersion relation, the continuum dispersion relation yields a fit with $\chi^2/\mathrm{d.o.f}= 4.90$ and results in:
\begin{eqnarray}
\label{g}
g_{\rho\pi\pi}&=&5.61(12)\FC\\
m_\rho\,a&=&0.4846(37)\FC \quad m_\rho=772(6)(8)\;\mathrm{MeV}\FD\nonumber
\end{eqnarray}
These results agree within error with the results based on the continuum dispersion relation,  that were presented in the paragraph after Eq. (41). 

\end{document}